\documentclass[intlimits,twoside,a4paper]{article}

\usepackage{amsmath,amssymb}
\usepackage{graphicx}

\usepackage[T2A]{fontenc}
\usepackage[cp1251]{inputenc}

\usepackage[eqsecnum]{cmpj2}

\usepackage{color}

\issue{2014}{17}{3}{33002}
\doinumber{10.5488/CMP.17.33002}

\newcommand{\eq}[1]{(\ref{#1})}
\newcommand{\fig}[1]{figure~\ref{#1}}

\newcommand{\be}{\begin{equation}}
\newcommand{\ee}{\end{equation}}

\newcommand{\la}{\left<}
\newcommand{\ra}{\right>}

\title[Lamplighter model of a copolymer adsorption]
{Lamplighter model of a random copolymer adsorption on a line}
\author[L.I. Nazarov, S.K. Nechaev, M.V. Tamm]
{L.I. Nazarov\refaddr{1}, S.K. Nechaev\refaddr{2,3,4}, M.V. Tamm\refaddr{1,4}}
\addresses{\addr{1} Physics Department, M.V. Lomonosov Moscow State University, 119991 Moscow,
Russia \addr{2} Universit\'e Paris-Sud/CNRS, LPTMS, UMR8626, B\^at. 100, 91405 Orsay, France
\addr{3} P.N. Lebedev Physical Institute, RAS, 119991 Moscow, Russia \addr{4} Department of Applied
Mathematics, National Research University ``Higher School of Economics'',  \\ 101000 Moscow, Russia}

\date{Received April 30, 2014}
\authorcopyright{L.I. Nazarov, S.K. Nechaev, M.V. Tamm, 2014}

\begin{document}

\maketitle

\begin{abstract}
We present a model of an AB-diblock random copolymer sequential self-packaging with local quenched
interactions on a one-dimensional infinite sticky substrate. It is assumed that the A-A and B-B
contacts are favorable, while A-B are not. The position of a newly added monomer is selected in view of
the local contact energy minimization. The model demonstrates a self-organization behavior with the
nontrivial dependence of the total energy, $E$ (the number of unfavorable contacts), on the number
of chain monomers, $N$: $E\sim N^{3/4}$ for quenched random equally probable distribution of A- and
B-monomers along the chain. The model is treated by mapping it onto the ``lamplighter'' random walk
and the diffusion-controlled chemical reaction of  $X+X\to 0$ type with the subdiffusive motion of
reagents.
\pacs 05.40.-a, 02.50.Ga, 87.15.Cc
\end{abstract}


\section{Introduction}

In this letter we propose a simple one-dimensional model of stochastic dynamic system possessing
a local optimization in quenched environment. The question of choosing an optimal strategy for a dynamic
system if only \textit{local} optimization is accessible, is a generic problem far
beyond the scope of natural science. Let us imagine that someone knows nothing about the future
and adapts his or her own behavior in each current time moment only on the basis of the knowledge about
the best solution at a given narrow time slice. Such a behavior leads to an optimal local strategy
when only a partial (current) knowledge is accessible, though it might be far from generic optimal
one if the knowledge about the future is available.

As an example of physical problems sharing the properties of local and global optimization, we can mention the
problems of protein secondary structure formation and DNA packaging in a viral capsid. If
in the course of protein folding, the chain forms a ``frozen'' network of contacts, then the optimal
conformation is determined only by a sequential step-by-step optimization of the added heteropolymer
fragments. However, if the whole protein is capable of ``adjusting'' the structure of a contact
network by breaking some bonds and creating other bonds, then the global optimization might essentially
change the folding picture.

The system considered below should not be regarded as a model of any specific physical system
(though it has some features of a protein folding), but it rather highlights the principles of
local optimization of a specific one-dimensional stochastic system in quenched random environment,
which could lead to a nontrivial self-organization.

Our model resembles the ``lamplighter random walk'' proposed in 1973 by A.M.~Vershik in the afterword
to the Russian edition \cite{greenleaf} of the book ``Invariant means on topological groups and
their applications'' by F.P. Greenleaf as an example of nontrivial estimate for the growth of
various numerical characteristics of groups. The object known as a ``lamplighter group'' (the name
was also given by A.Vershik) became very popular among probabilists after the pioneering work by
A.~Vershik and V.~Kaimanovich \cite{kaiman}.

\section{The model}

\begin{figure}[!h]
\centerline{
\includegraphics[width=8cm]{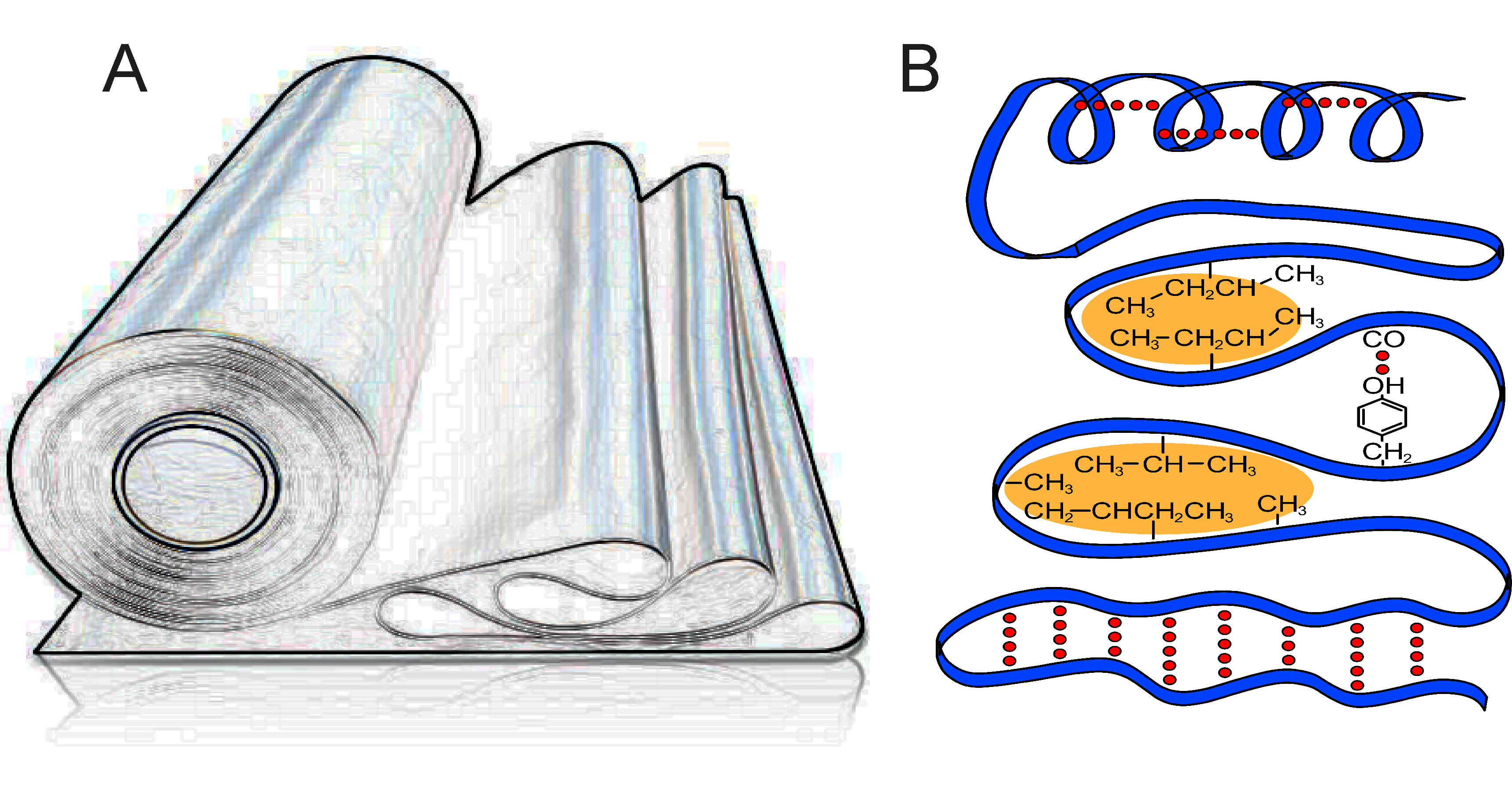}
}
\caption{(Color online) (a) Folds of tissue getting out of the roll; (b) Secondary structure formation in a
polypeptide heteropolymer chain.}
\label{fig:01}
\end{figure}

The toy model presented below imitates the packaging of a random diblock copolymer with quenched
sequence of monomers on a one-dimensional substrate. Schematically it can be viewed as a tissue
folding getting out of the roll, as depicted in  \fig{fig:01}~(a), where the size of folds depends
on local heterogenous interactions between different parts of the tissue. Such a schematic view
might be compared with the specific secondary structure formation of a linear polypeptide chain
in \fig{fig:01}~(b), where two main types of secondary structure, $\alpha$-helices and $\beta$-sheets,
are shown and the secondary structure is stabilized by the hydrogen bonds.

To pass from these intuitive pictures to a more specific description, consider a $N$-monomer chain
with quenched random primary sequence of two types of links, A and B, placed on an infinite line,
which is sticky for links A and repulsive for links B. We suppose that the energy of the A-A or B-B
contacts is $0$, while the energy of the A-B contact is $+1$ (in dimensionless units), i.e., the
monomers of the same types are indifferent to each other, while the monomers of different types
repulse.

\begin{figure}[!b]
\centerline{
\includegraphics[width=8cm]{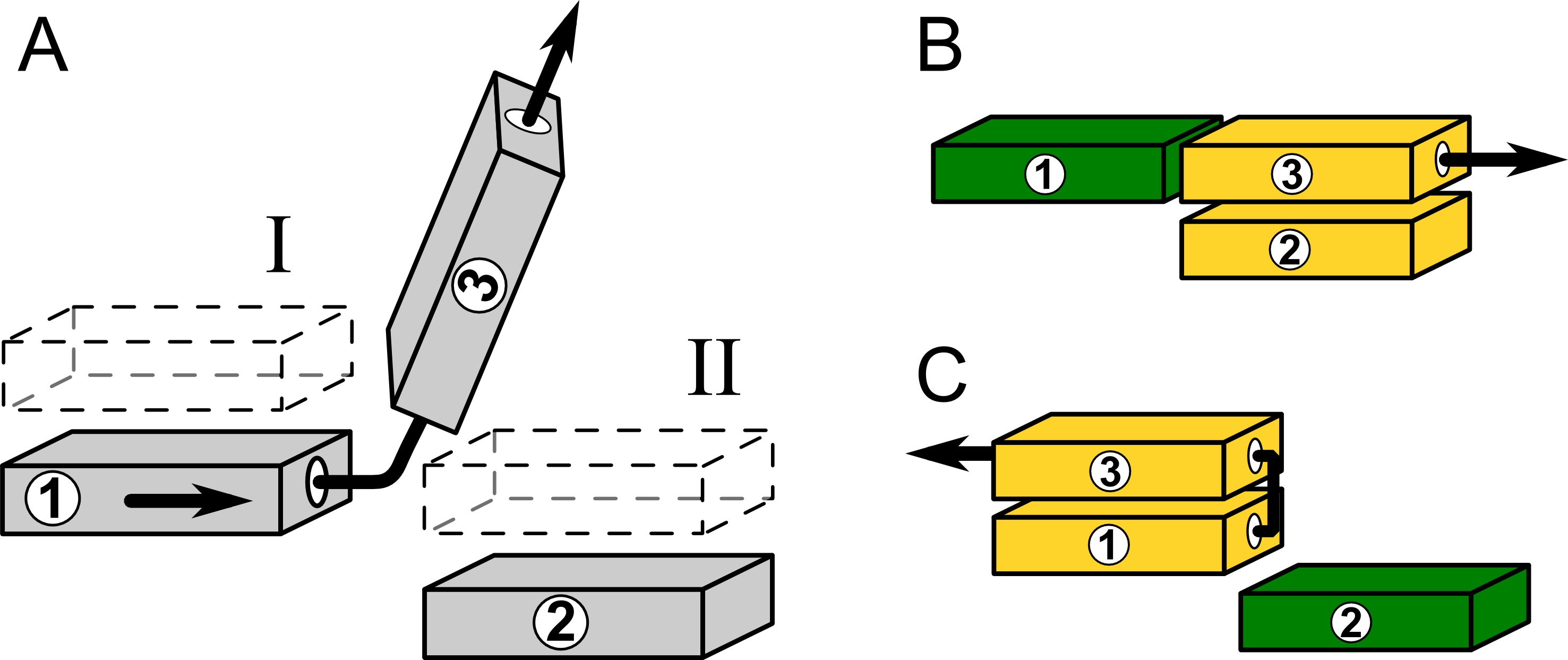}
}
\caption{(Color online) Local rules of the monomers freezing to the structure. (a) The monomer 3 has two possible
locations I and II. The selection of the position depends on the types of all three monomers. (b)
Monomers 2 and 3 are identical, while the monomer 1 differs from 2 and 3. (c) The inverse
situation with respect to~(b).}
\label{fig:02}
\end{figure}

The heteropolymer folding (packaging) on an infinite line is an irreversible sequential process of
adding monomers to the already existing frozen environment. Let us recursively describe the folding process. For definiteness,
suppose that the line is initially uniformly covered with monomers of
type A and let the first chain monomer be always of type A. Put the 1st monomer at a position
$x_{0}=0$ on the line. The 2nd monomer could be placed either next to the 1st monomer on a line in
the \emph{trans} state, or it can be put on top of the 1st monomer in the \emph{gauche} state,
making a ``hook'' as it is shown in \fig{fig:02}. The newly added monomer interacts with the one
located below it in the projection to the line.

\begin{figure}[!t]
\centerline{
\includegraphics[width=8cm]{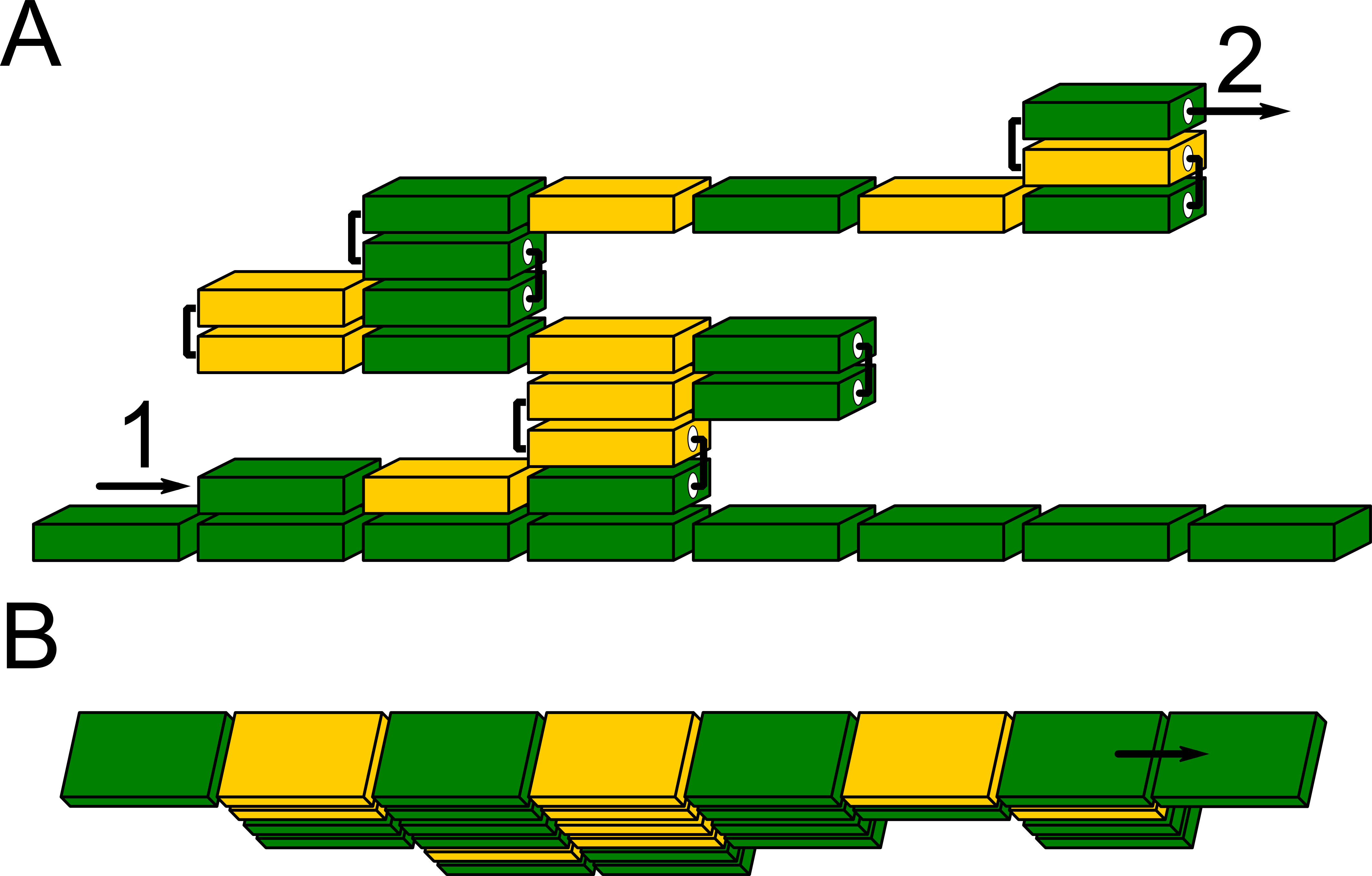}
}
\caption{(Color online) The 20-step of block copolymer package on the plate with the uniform monomer type
distribution [(a) The side view; (b) The top view].}
\label{fig:03}
\end{figure}

The selection between \emph{trans-} and \emph{gauche-} states is made according to the following
rules: i) if one of two possible new (\emph{trans-} or \emph{gauche-}) states is favorable and the
other is not, then the favorable state is always selected; ii) if new \emph{trans-} and
\emph{gauche-} states are both favorable or both unfavorable, then a new conformation
(\emph{trans-} or \emph{gauche-}) is chosen randomly with the probability $1/2$. For any ``defect''
(i.e., unfavorable pairing) the energy penalty $+1$ is added. The sample structure of the first 20 steps
is depicted in \fig{fig:03}.

The position of a newly added monomer is selected in view of the irreversible local contact energy
minimization, while for equal energies, the choice of \emph{trans-} or \emph{gauche-} states is
random with equal probabilities. Let us emphasize that only the uppermost monomers or the ``roof'' of
the structure [i.e., the ones visible from the top of the picture in \fig{fig:03}~(b)] take part in
the play: the monomers that are underneath do not participate in a further optimization process.
Therefore, the folding is an effective Markovian process, whose states
are specified by i)  the set of monomers visible currently from the top of the structure (i.e., by the ``roof''), and ii) the position and
direction of the last step, which defines the positions where the next monomer can be
added.

\section{Results}

\begin{figure}[!t]
\centerline{
\includegraphics[width=0.32\textwidth]{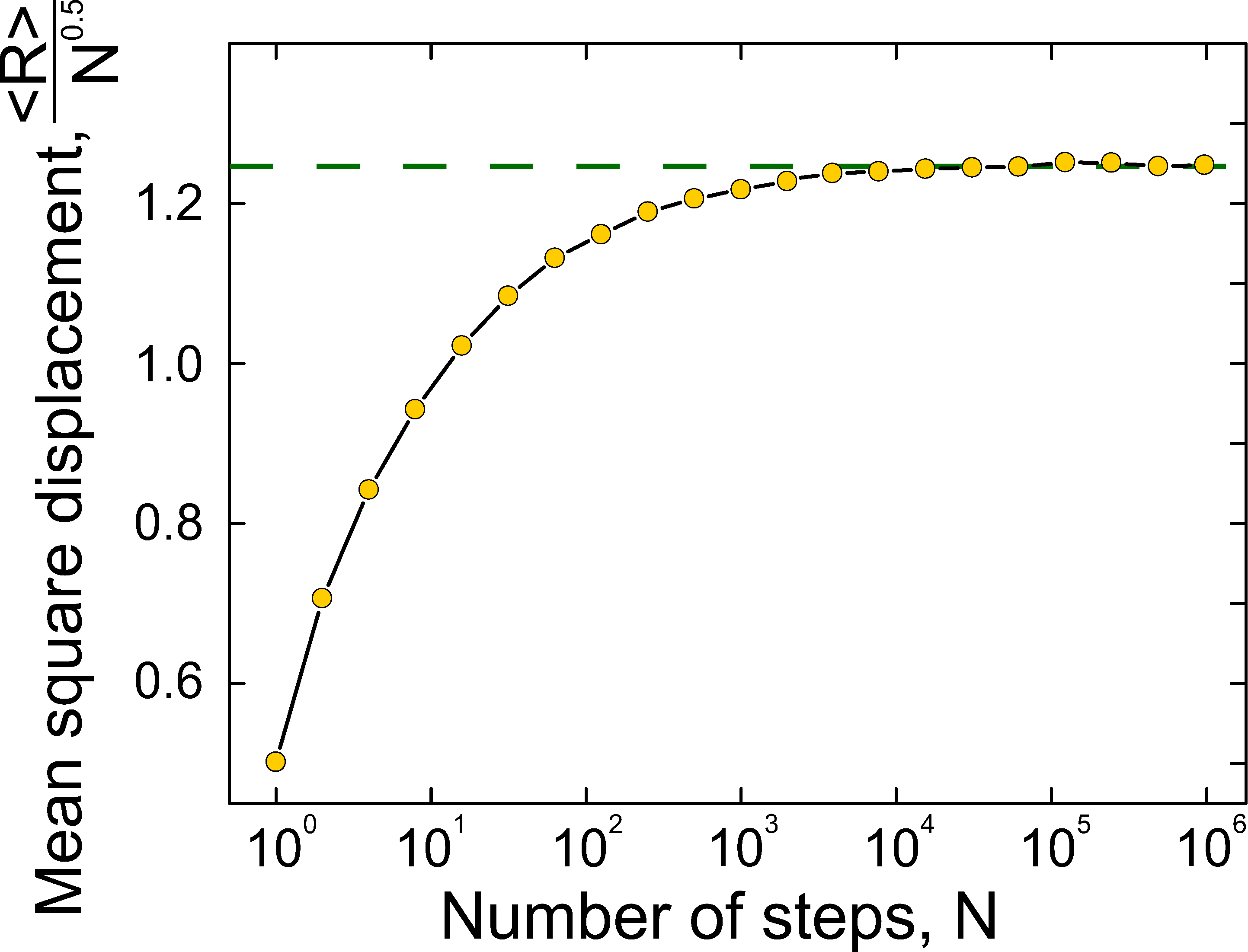}
\includegraphics[width=0.33\textwidth]{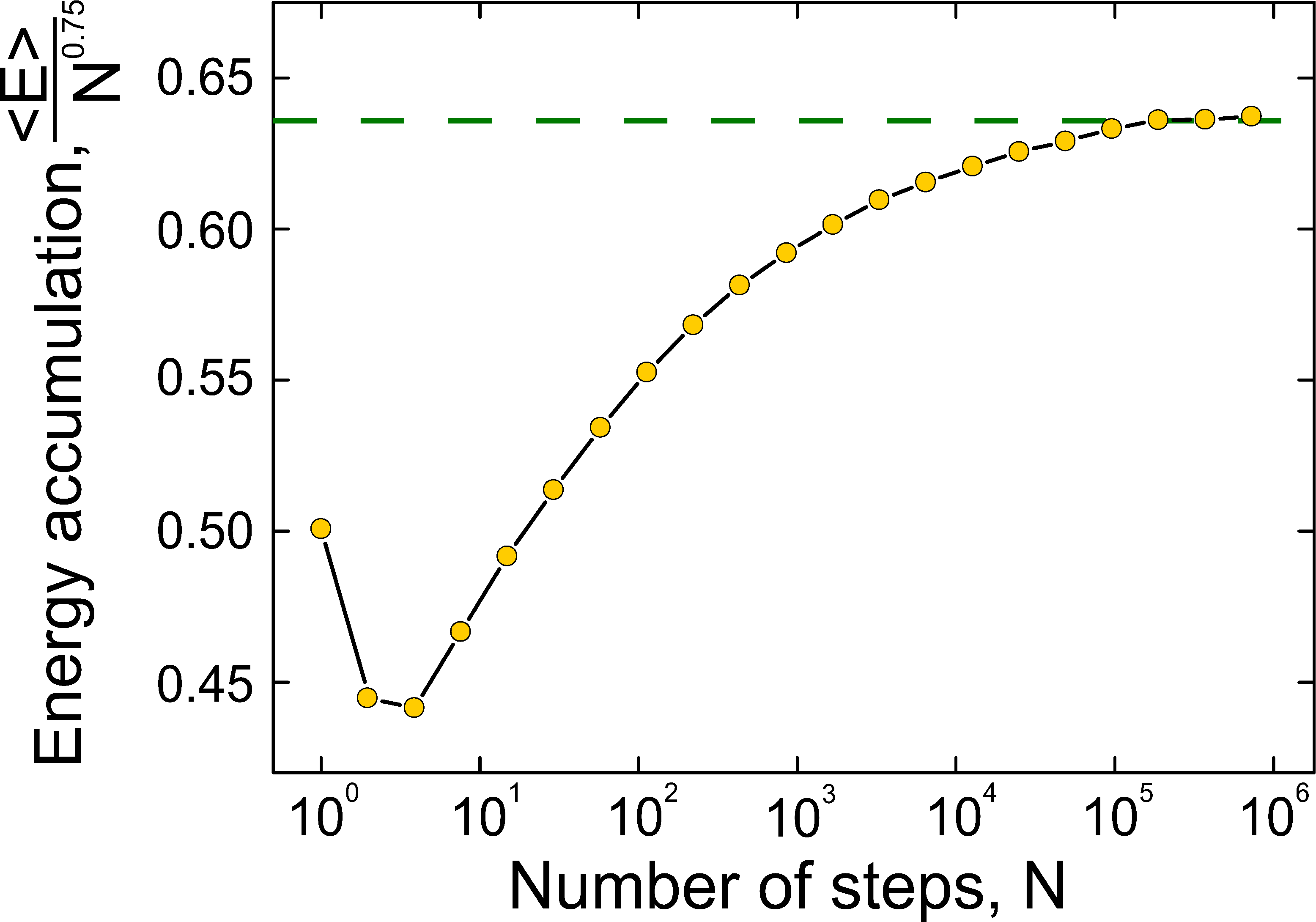}
\includegraphics[width=0.33\textwidth]{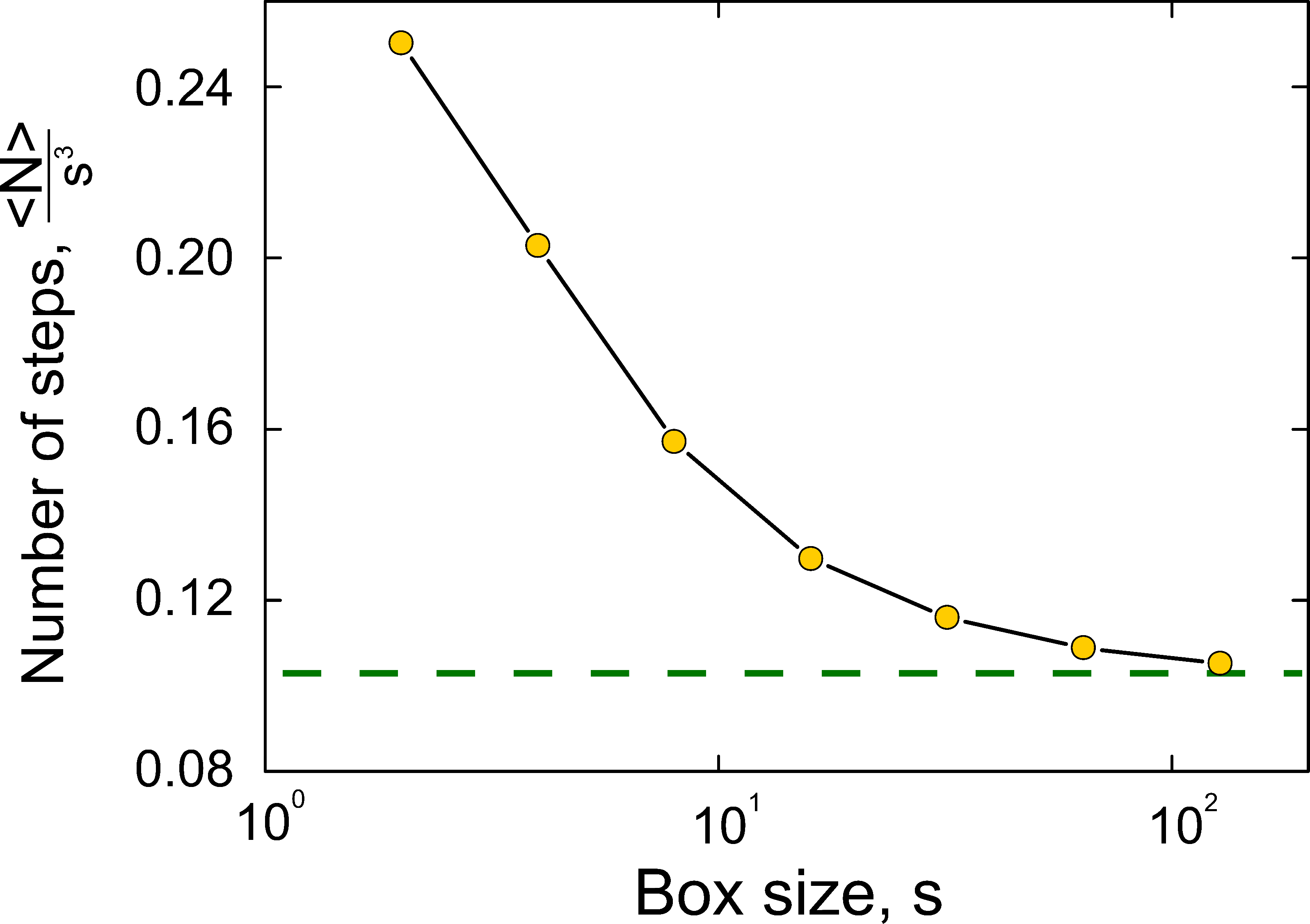}
}
\caption{(Color online) (left): The average normalized span, $\la R^2(N) \ra$, of the folded trajectory on an
infinite substrate, as a function of the number of steps, $N$; (center): The normalized cumulative
folding energy, $E(N)/N^{3/4}$, as a function of the number of steps, $N$; (right): The typical
normalized demixing time, $\tau/s^3$, of the folded trajectory as a function of the bounding box
size, $s$.}
\label{fig:06}
\end{figure}

In this section we consider the properties of the described locally optimal folding averaged over a
set of completely random initial primary sequences of a polymer, i.e., we assume A and B monomers
to appear in the chain with equal probabilities without any correlations between sequential
letters. The most natural question to ask about this folding is how the average total energy
penalty, $\la E\ra$, depends on the chain length, $N$, i.e., how many unfavorable A-B contacts one
has on average in the $N$-monomer chain. Unexpectedly, it turns out that $\la E\ra$ grows
sub-linearly with $N$, namely as $\la E\ra \sim N^{3/4}$ [see \fig{fig:06}~(center)].

\begin{figure}[!t]
\centerline{
\includegraphics[width=10cm]{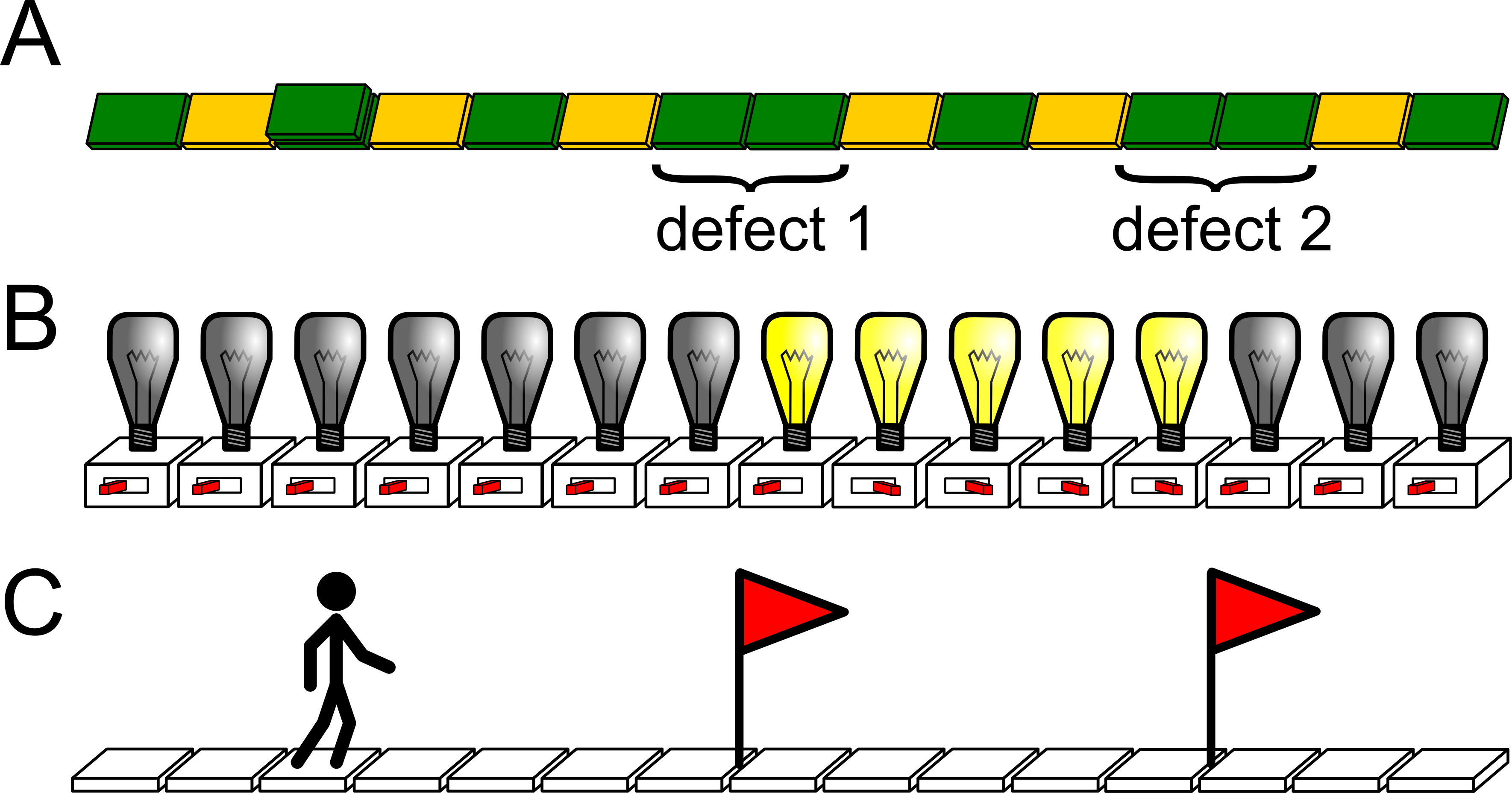}
}
\caption{(Color online) The representation of two arbitrary packages in a context of 1D walks on an adjustable
substrate. Periodic regions, shown in (a) are mapped (depending on the parity) to the domains of
lamps in ``off'' or ``on'' positions, shown in (b). The white segments, shown in (c), correspond to
the domains of lamps, while red flags mark the domain walls. The human figure designates the tip of the
polymer chain, which performs a random walk and can push flags to the
left or to the right when passing by them.}
\label{fig:04}
\end{figure}

To understand this result, one should notice that in terms of the evolution of the roof of the
structure defined above, the system has an adsorbing state. Indeed, if the roof is represented by
an alternating sequence, i.e., \ldots-A-B-A-B-\ldots, every newly added monomer can be positioned on the
top of this periodic sequence \emph{without any energetic penalty}, and the resulting structure of
the roof will stay unchanged. Moreover, if some part of the roof is alternating [as in
\fig{fig:03}~(b)], this part will remain alternating and any sequence can be adsorbed on it without
penalty. Thus, the heteropolymer folding in our model can be considered as a dynamic process of
approaching the adsorbing alternating state. This dynamics consists in sequential cancellation of
``defects'', represented by non-alternating pairs of A-A or B-B of consecutive monomers on the
substrate as it is shown in \fig{fig:04}~(a).

To formalize this idea, we introduce an Ising variable, $s(x)$, which labels the state of the
roof in position $x$:
\be
s(x)= \left\{
\begin{array}{ll}
+1 & \text{for monomer A in position $x$},  \\
-1 & \text{for monomer B in position $x$}.
\end{array} \right.
\label{Ising1}
\ee
Now, we define another Ising variable, $\sigma(x)$:
\be
\sigma(x) = (-1)^x s(x).
\label{Ising2}
\ee
The change of variables form $s(x)$ to $\sigma(x)$ essentially simplifies the picture, because in
terms of the variable, $\sigma$, the adsorbing state is homogenous. The state of the chain in terms
of variables $\sigma$ is depicted in \fig{fig:04}~(b) by lamps switched on (if $\sigma = 1$) and off
(if $\sigma = -1$). Each defect (i.e., the pair of segments A-A or B-B) is a ``domain wall''
separating two states: to the left of it, the lamps are off, while to the right of it, the lamps are
on. A configuration with two defects is depicted in \fig{fig:04}~(b). The defects (domain walls)
are shown in \fig{fig:04}~(c) by flags. In terms of variables $\sigma$, the uniform adsorbing state
corresponds to the periodic adsorbing state of our initial Ising model, and the dynamics of folding
in terms of $\sigma$ means coarsening of the Ising domains.

We can easily formulate the dynamic rules describing the evolution of the system. The tip of the
original polymer chain, shown as a human figure (``lamplighter'') in  \fig{fig:04}~(c) performs a walk on a
one-dimensional lattice. In the absence of any energy gain, the \emph{trans}-- and
\emph{gauche}--states are equiprobable. Since the concentrations of monomers of types A and B in
the chain are the same, the lamplighter performs a symmetric random walk on a one-dimensional discrete
line as it is shown in \fig{fig:04}~(c).

The initial state for the spins $\sigma$ is an infinite one-dimensional lattice with all domains of
length 1. Using the notations of \fig{fig:04}~(c), the flags are located at each boundary between
adjacent monomers, while a single random walker (lamplighter) is initially positioned at $x=0$. As the
time (measured in the number of heteropolymer monomers, $N$) runs, the lamplighter performs a
discrete symmetric random walk with the following properties: i) if the lamplighter stays within a
domain of lamps which all are ``on'' or ``off'', he does nothing with the lamps, ii) if the lamplighter
crosses the domain wall, marked by a flag, he may (with the probability 1/2) randomly shift the flag on one
lattice site to the left or to the right. If two flags meet each other at one site, they
annihilate. The energy of the folding configuration is equal to the number of flag shifts. As the
flags annihilate and their concentration decreases, the energy of the folding grows slower.

To consider the dynamics more quantitatively, we have first checked our conjectures computing
numerically the span, $\la R^2(N) \ra$, of the lamplighter's $N$-step walk. As shown in
\fig{fig:06}~(left), we have got
\be
\la R^2(N) \ra \sim N
\label{eq:05}
\ee
as expected for the standard Brownian random walk. In terms of the original problem, this means that
the $N$-step random block-copolymer has a typical span of size of order $\sqrt{N}$ on the 1D
substrate.

Now, let us pay attention to the dynamics of a single flag. We have already mentioned that the
lamplighter passing across the flag, moves it on one lattice site to the left or to the right with
equal probabilities. This means that the flag itself performs a random walk with the mean-square
displacement $\la\Delta x^2(s)\ra \sim m$, where $m$ is the number of jumps the flag has made (i.e.,
the number of times the lamplighter returned to the flag). Since the number of returns to a given point
(the position of a flag), $m$, scales as $m \sim t^{1/2} $ for a one-dimensional random walk at
large $t$, we immediately conclude that the  mean-square displacement of the flag obeys the following
scaling dependence
\be
\la \Delta x^2(t) \ra \sim m \sim t^{1/2}.
\label{eq:01}
\ee

The waiting times between sequential visits of a given flag by the lamplighter are distributed as
the first-return times of a one-dimensional random walk, with a tail $t^{-3/2}$ at long times. This makes the movement
of a single flag exactly the continuous-time random walk (CTRW) with the
exponent $1/2$ \cite{ctrw}. Note that since the movement of different flags is governed by a
\emph{single} lamplighter, they are correlated. However, in the first approximation one expects
that it is possible to neglect these correlations and assume that the mean square distance,
$d=\sqrt{\la(x_{i+1}-x_i)^2\ra}$ between two neighboring flags $i$ and $i+1$ located at the points
$x_i$ and $x_{i+1}$, behaves as in \eq{eq:01}, i.e. $d^2(t)\sim t^{1/2}$.

The process of annihilation of flags is described by the equation of a one-dimensional chemical
kinetics $X+X\to 0$, where the reagent $X$ designates a flag, and $X$ experiences a sub-diffusive
random walk at time $t$ with the mean-square displacement $\la\Delta x^2(t)\ra \sim t^{1/2}$. Such
a process has been widely studied in chemical kinetics~--- see, for example, \cite{lind}.

Let $c(t)$ be the typical concentration of $X$ at time $t$ within a region visited by the
lamplighter. We are interested in the survival probability of $X$ in time, i.e., in the scaling
behavior $c(t)\sim t^{-\alpha}$ at $t\to\infty$. Each time the two consecutive flags touch each
other, they disappear. Since the flags cannot pass through each other, the typical distance between
the reagents $X$ (flags) at time $t$ is of the order of the explored territory up to the time $t$.
Therefore, $c(t) \sim d^{-1}(t)$ or
\be
c(t) \sim t^{-1/4}
\label{eq:02}
\ee
giving $\alpha = 1/4$.

The energy of the folded chain increases by one with the probability $1/2$ every time the
lamplighter visits a flagged site. There are two typical situations when it happens: a) when the
lamplighter reaches the boundary of the  domain visited (recall that outside the  domain visited all
sites are flagged), and b) when it encounters a flag in the bulk of the already visited region. It is
easy to estimate the frequencies of both these events. Indeed, according to \eq{eq:05}, the size of
the domain visited grows as $t^{1/2}$, and the resulting energy increment scales as
\be
\left.\frac{\rd E(t)}{\rd t}\right|_\textrm{boundary} \sim \frac{\rd t^{1/2}}{\rd t} \sim t^{-1/2}.
\label{eq:03a}
\ee
On the other hand, the probability for a random walk to meet a flag at time $t$ in the domain's
bulk, scales as $c(t)$ and, therefore,
\be
\left.\frac{\rd E(t)}{\rd t}\right|_\textrm{bulk} \sim d^{-1}(t) \sim t^{-1/4}.
\label{eq:03}
\ee
We see that, at least on large timescales (i.e., for long chains), the second mechanism dominates,
and the energy is mostly accumulated via interaction of the lamplighter with the flags in the bulk. Integrating \eq{eq:03}
up to the total chain length, $N$, we end up with the scaling behavior of the average energy
accumulation, $E(N)$ up to the length $N$ ($N\gg 1$):
\be
\la E(N) \ra \sim \int_{0}^N t^{-1/4}\, \rd t \sim N^{3/4}.
\label{eq:04}
\ee
This scaling behavior is clearly seen in the numeric simulations shown in \fig{fig:06}~(center),
where we have plotted the normalized cumulative energy, $E(N)/N^{3/4}$, as a function of $N$. The
results are obtained by averaging over $10^6$ random block-copolymer packaging attempts on the
infinite line.

Apart from the chain adsorption on an \emph{infinite} one-dimensional substrate, it seems
instructive to consider the case when the total size of a substrate is \emph{finite}, e.g., the
chain is trapped in a cage with periodic boundary conditions. In this case (for even total number
of sites, $L$, in the cage),  a long enough chain, once again, finally forms a
state with completely regular \ldots A-B-A-B \ldots roof, which means, in terms of the Ising domain representation, that a single
Ising domain is formed. The natural questions to ask are as follows: i) how long (on average) does
it take for a perfect domain structure to be formed in the cage of $L$ sites (i.e., which is the
typical ``demixing time'', $\tau$, as a function of $L$), and ii) what is the typical total energy,
$\la E(L) \ra$ (the number of defects in the structure), as a function of the cage size, $L$. It turns out
that the question ii) is easier to answer. Indeed, the energy $E$ is the total number of the flag
moves, regardless of the waiting times between them. Since the moves of the flags in both
directions are equiprobable, the typical number of the moves needed for two adjacent flags to
annihilate, scales as $d^2$, where $d$ is the initial distance between them. Therefore, one expects
the total accumulated energy to scale as
\be
\la E(L) \ra \sim \sum_{n=0}^N d^2(n),
\label{eq:07}
\ee
where $n$ ($1\leqslant n \leqslant L$) is the number of surviving flags at a given time (all numeric constants
are omitted). Since the typical distance between the flags is $d(n) \sim L/n$, we can evaluate the sum
in \eq{eq:07}, getting finally
\be
\la E(L) \ra \sim L^2
\label{eq:07a}
\ee
for large $L$, which is perfectly confirmed by numeric simulations.

For the demixing time, we have not yet obtained the analytic scaling estimate, but according to the
numeric simulations shown in \fig{fig:06}~(right), the demixing time $\tau (L)$ scales as
\be
\tau(L) \sim L^3.
\label{eq:06}
\ee
Combining \eq{eq:07a} and \eq{eq:06}, we get a nontrivial prediction $\la E(\tau) \ra \sim
\tau^{2/3}$, different from the result of the mean-field theory \eq{eq:04} in the infinite domain
($\tau$ plays the role of the chain length). Therefore, it turns out that in a bounded domain, the
shifts of sequential flags are highly correlated since they are governed by a single lamplighter
which cannot escape the cage. Indeed, when only several flags are left in a
cage, the correlations between their shifts become very important. A more detailed consideration of
this phenomenon will be provided elsewhere.

\section{Discussion}

We have  already mentioned that our model is reminiscent to the so-called ``lamplighter random walk'',
widely considered in the mathematical literature (see, for example,
\cite{greenleaf,kaiman,lyons,grig,kaiman2,rev1,rev2,peres,woess}). Here is the definition of the
\emph{lamplighter graph}, taken from \cite{woess}: ``\ldots Think of a (typically infinite) connected
graph $X$, where in each vertex there is a lamp that may be switched off (state 0), or switched on
with $q-1$ different intensities (states $1,\ldots, q-1$). Initially, all lamps are turned off, and a
lamplighter starts at some vertex of $X$ and walks around. When he visits a vertex, he may switch
the lamp sitting there into one of its $q$ different states (including ``off''). Our information
consists of the position $x\in X$ of the lamplighter and of the finitely supported configuration
$\eta: X\to \mathbb{Z}_q=\{0,\ldots, q-1\}$ of the lamps that are switched on, including their
respective intensities. The set $\mathbb{Z}_q \wr X$ of all such pairs $(\eta,x)$ can be
in several ways equipped with a naturally connected graph structure, giving rise to a \emph{lamplighter
graph}\ldots''

The key difference between the lamplighter model considered in the mathematical literature
(math-LL) and the model analyzed in our work, consists in the following: in math-LL all lamps
(governed by a single random walk) are immobile, are located at every lattice site, and do not
interact with each other, while in our model,  the flags (also governed by a single random walk) can
perform random local shifts and interact (i.e., annihilate while in a contact). Despite the mentioned
difference, we believe that some important questions, such as the computation of the
``mixing time'' \cite{peres} on lamplighter graphs, could be addressed in our model as well.

It seems interesting to compare the results of the local step-by-step optimization discussed at length in this work with the global optimization of the chain configuration. Namely, assume that
after the described packaging, the frozen configuration of folds is annealed, i.e., the existing
folds are allowed to retract and the new folds can be formed (for unchanged sequence of monomers).
Clearly, the resulting ground state energy will be lower than the one obtained in our local
optimization process. The partition function of such a system having an  annealed secondary structure,
apparently admits the recursive representation similar to the dynamic programming algorithm for RNA
secondary structure \cite{valba}. In the forthcoming works we plan to provide a detailed analysis
of the global optimization and its comparison with the results of local optimization for the
lamplighter random walk.

\section*{Aknowledgements}

The authors are grateful to V.~Avetisov, P.~Krapivsky and R.~Metzler for encouraging discussions.
This work was partially supported by the grants ANR-2011-BS04-013-01 WALKMAT and the IRSES project
FP7-PEOPLE-2010-IRSES 269139 DCP-PhysBio. M.T. and S.N. acknowledge the financial support of the
Higher School of Economics program for Basic Research.

\clearpage

\ukrainianpart

 \title{Модель ліхтарника адсорбції випадкового кополімера на лінії}

 \author{Л.І. Назаров\refaddr{1}, С.К. Нєчаєв\refaddr{2,3,4}, М.В. Тамм\refaddr{1,4}}
 \addresses{\addr{1} Фізичний факультет, Московський державний університет ім. М.В. Ломоносова, Москва, РФ
 \addr{2} Університет Парі-Сюд/CNRS, Орсе, Франція
 \addr{3} Фізичний інститут ім. П.Н. Лєбєдєва, РАН, Москва, РФ
 \addr{4} Факультет прикладної математики, Національний дослідницький університет
 ``Вища школа економіки'', Москва, РФ}

 \makeukrtitle

 \begin{abstract}
 \tolerance=3000%
 Ми представляємо модель послідовного самопакування AB-диблокового випадкового кополімера з локальними замороженими взаємодіями
 на одновимірній нескінченій липкій основі. Припускається, що контакти A-A і B-B є сприятливі, тоді як A-B є несприятливими.
 Положення нового мономера, що додається, виберається з точки зору мінімізації енергії локального контакту.
 Модель демонструє саморганізовану поведінку з нетривіальною залежністю загальної енергії, $E$ (числа несприятливих контактів), від
 числа мономерів ланцюга, $N$: $E\sim N^{3/4}$ для замороженого хаотичного рівноймовірного розподілу A- і
 B-мономерів вздовж ланцюга.
 Модель розглядається шляхом зіставлення її з
 випадковим блуканням ліхтарника і дифузійно-контрольованою хімічною реакцією типу $X+X\to 0$ з субдифузійним рухом реагентів.
 \keywords локальна оптимізіція, згортання гетерополімера, випадкове блукання ліхтарника, субдифузійна хімічна реакція
 \end{abstract}

\end{document}